\documentclass[conference]{IEEEtran}

\IEEEoverridecommandlockouts
\usepackage{cite}
\usepackage{amsmath,amssymb,amsfonts}
\usepackage{algorithmic}
\usepackage{graphicx}
\usepackage{textcomp}
\usepackage{xcolor}
\usepackage{psfrag}
\usepackage{tikz}
\usepackage{multicol}
\usepackage{cuted}
\usepackage{epsfig}
\usepackage{color}
\usepackage{geometry}
\def\BibTeX{{\rm B\kern-.05em{\sc i\kern-.025em b}\kern-.08em
    T\kern-.1667em\lower.7ex\hbox{E}\kern-.125emX}}

\newcommand{\bm}{\mathbf}
\newcommand{\bs}{\boldsymbol}

\begin{document}

\title{Delay-Doppler Multiplexing with Global Filtering \vspace{-0.1 cm}}
\author{\IEEEauthorblockN{Jialiang Zhu, Mohsen Bayat, Arman Farhang
} 
\IEEEauthorblockA{Department of Electronic \& Electrical Engineering, Trinity College Dublin, Ireland \\ \{zhuj3, bayatm, arman.farhang\}@tcd.ie}
\vspace{-0.6 cm}
}


\maketitle

\begin{abstract}
    This paper proposes a novel modulation technique called globally filtered orthogonal time frequency space (GF-OTFS) which integrates single-carrier frequency division multiple access (SC-FDMA)-based delay-Doppler representation with universal filtered multi-carrier (UFMC) modulation. 
    Our proposed technique first arranges the frequency-Doppler bins of an orthogonal time frequency space (OTFS) frame in adjacency using  SC-FDMA 
    and then applies universal filtering to the neighboring signals to mitigate inter-Doppler interference (IDI).
    By employing this approach, GF-OTFS achieves superior spectral containment and effectively mitigates interference caused by Doppler shifts in dynamic, time-varying channels.
    This paper also presents a detailed mathematical formulation of the proposed modulation technique. Furthermore, a comprehensive performance evaluation is conducted, comparing our GF-OTFS approach to state-of-the-art techniques, including Doppler-resilient UFMC (DR-UFMC) and receiver windowed OTFS (RW-OTFS). Key performance metrics, such as bit error rate (BER) and out-of-band (OOB) emissions, as well as the Doppler spread reduction are analyzed to assess the effectiveness of each approach. The results indicate that our proposed technique achieves comparable BER performance while significantly improving spectral containment. 
\end{abstract}


\vspace{-0.6cm}
\section{Introduction}
\vspace{-0.0cm}

Reliable communication for high-mobility users is a key challenge in next-generation wireless networks. The combination of high mobility and the demand for high data rates leads to doubly-selective wireless channels. 
While orthogonal frequency division multiplexing (OFDM) is widely used in modern wireless communication systems due to its robustness against multipath fading, its vulnerability to Doppler spread in time-varying environments remains a significant drawback \cite{b1}.
Doppler spread leads to frequency shifts, which break the orthogonality of OFDM subcarriers and leads to performance degradation.
To address this issue, orthogonal time frequency space (OTFS) modulation was introduced in \cite{b2}.
OTFS places the data symbols on the delay-Doppler plane rather than the time-frequency plane.
This provides inherent resilience to the Doppler effect and makes OTFS a strong candidate for deployment in future networks \cite{b2, b3, b4}. 

Despite the clear advantages of OTFS that are well articulated in \cite{b2}, achieving superior performance in terms of spectral efficiency and mitigating Doppler effects under time-varying channel remains a significant challenge.
To address this issue, 
three groups of solutions have been proposed. 
The first group includes Zak OTFS and orthogonal delay-Doppler multiplexing (ODDM), which directly map data symbols onto orthogonal basis functions in the delay-Doppler domain \cite{b5,b3}. These techniques enhance spectral efficiency while preserving OTFS's robust performance in managing high mobility and Doppler effects.
In the second group of solutions,
OTFS has been combined with various time-frequency modulation techniques, including generalized frequency division multiplexing (GFDM) \cite{b31}, filter bank multicarrier (FBMC) \cite{b32}, and universal filtered multi-carrier (UFMC) \cite{b7}.
Among these efforts, a notable example is the Doppler-resilient UFMC (DR-UFMC), presented in \cite{b7}, which applies UFMC on top of OTFS modulation for time-frequency transmission.
In this technique, UFMC is applied to each delay block, which increases the block length, 
followed by the overlap-and-add process to ensure the transmitted signal length remains comparable to that of a full cyclic prefix (CP) OTFS.
As shown in \cite{b6}, most works in the previously mentioned two groups concentrate on pulse shaping along the delay domain. Hence, while these techniques enhance spectral containment, they do not fully utilize the filtering properties along the Doppler dimension to mitigate inter-Doppler interference (IDI).
To address the issue of fractional Doppler effects, a third group of modulation techniques has been proposed. 
In this group, the authors of \cite{b8} proposed receiver windowed OTFS (RW-OTFS), which applies a receiver window to reduce Doppler-induced leakage. Although this technique effectively mitigates interference caused by fractional Doppler shifts, it does not reduce out-of-band (OOB) emissions, leaving spectral efficiency largely unaddressed.
Recently, a single-carrier frequency division multiple access (SC-FDMA)-based delay-Doppler representation has been presented to achieve performance equivalent to that of the original OTFS, as demonstrated in \cite{b9}.
In this implementation, delay-Doppler symbols are first transformed into the frequency-Doppler domain. Then, a one-dimensional inverse discrete Fourier transform (IDFT) is applied across the entire frequency-Doppler signal samples to generate the time-domain signal.

By examining the OTFS signal from SC-FDMA implementation perspective, in this paper, we propose a novel modulation technique called globally filtered OTFS (GF-OTFS).
This technique integrates UFMC \cite{b11}, with SC-FDMA-based delay-Doppler signal representation. 
The proposed GF-OTFS arranges the frequency-Doppler bins of an OTFS frame adjacently using SC-FDMA and applies universal filtering to the neighboring signals to mitigate IDI.
Unlike DR-UFMC, our proposed technique employs UFMC to mitigate IDI by applying pulse shaping in the frequency-Doppler domain.
The proposed technique further enhances spectral containment and significantly reduces OOB emissions compared to OTFS, Zak-OTFS, ODDM, and RW-OTFS.
In this work, to evaluate the efficiency of the modulation techniques in mitigating Doppler-induced leakage, and we consider a scenario with ideal pulse shaping along the delay dimension.
Regarding the bit-error-rate (BER), our proposed technique does not reach an error floor compared to RW-OTFS.
It also achieves up to approximately $5~\rm{dB}$ gain at high signal-to-noise ratios (SNRs) compared to RW-OTFS and OTFS.

\textit{Notation:}
Scalars, vectors, and matrices are denoted by regular lowercase, bold lowercase, and bold uppercase letters, respectively.
The superscripts $(.)^{\rm{H}}$, $(.)^{\rm{T}}$, and $(.)^{\rm{*}}$ indicate Hermitian, transpose, and conjugate operations, respectively.
$\mathbb{C}$ represents complex number set.
The operators $\otimes$, $\odot$, and $\oslash$ denote Kronecker product, element-wise product, and element-wise division, respectively.
The function ${\rm{diag}} \{ \mathbf{x} \}$ constructs a diagonal matrix with the elements of the vector $\mathbf{x}$ on its main diagonal.
The notation $\bm{X} = \mathcal{T}_{M \times N} \{ \bm{x} \}$ for an $(M+N-1) \times 1$ vector $\bm{x}$, represents an $M \times N$
Toeplitz matrix, in which$[\bm{X}]_{mn} = [\bm{x}]_{m-n+N}$.
The vectors $\mathbf{1}_{ N}$ and $\mathbf{0}_{N}$ are all-one and all-zero column vectors of length $N$, respectively.
$\mathbf{I}_{N}$ is the identity matrix of size $N$, and $\mathbf{0}_{M\times N}$ and $\mathbf{1}_{M\times N}$ are $M\times N$ matrices consisting of zeros and ones, respectively.
$\mathbf{{F}}_{N}$ is the normalized $N$-point DFT matrix with the elements $[\mathbf{F}_{N}]_{m,n} =\frac{1}{\sqrt{N}} e^{-j \frac{2 \pi mn}{N}}$ for $m,n=0,\ldots, N-1$.

\vspace{-0.1cm}
\section{SC-FDMA-Based Implementation of OTFS}
\label{sec:model}
\begin{figure}[!t]
\psfrag{Ft}{\hspace{-0.2mm}\scriptsize{$\mathcal{F}_\tau$} }
\psfrag{Fv_t}{\hspace{-2mm}\scriptsize{$\mathcal{F}_\nu^{-1}$} }
\psfrag{Ff_t}{\hspace{-0.2mm}\scriptsize{$\mathcal{F}_{f}^{-1}$} }
\psfrag{F_1}{\hspace{-0.3  mm}\scriptsize{$\mathcal{F}_{f\nu}^{-1}$} }
\psfrag{Delay Doppler}{\hspace{-0.8  mm}\footnotesize{Delay-Doppler} }
\psfrag{Delay Time}{\hspace{-0.8  mm}\footnotesize{Delay-Time} }
\psfrag{Frequency Doppler}{\hspace{-0.8  mm}\footnotesize{Frequency-Doppler} }
\psfrag{Frequency Time}{\hspace{-0.8  mm}\footnotesize{Frequency-Time} }
\centering
\includegraphics[scale=0.216]{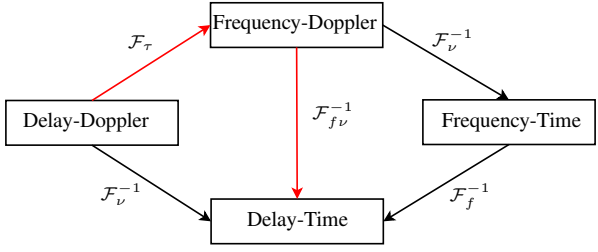} 
\vspace{-0.3 cm}
\caption{Relationships among the different domains.}
\vspace{-0.6 cm}
\label{fig:transform}
\end{figure}

\begin{figure}[!t]
\psfrag{F_M}{ \hspace{-1.4mm}$\mathbf{F}_M$}
\psfrag{F_MN}{ \hspace{-1.7 mm}$\mathbf{F}_{MN}^{\rm H}$}
\psfrag{M}{{$M$} }
\psfrag{x}{\hspace{-1.3 mm}{\small$\bm{x}_{\rm{t}}$} }
\psfrag{d0}{\hspace{-0.2mm}\scriptsize{$d_0[0]$} }
\psfrag{d1}{\hspace{-0.4mm}\scriptsize{$d_{M\!-\!1}[0]$} }
\psfrag{d4}{\hspace{-2.2mm}\scriptsize{$d_0[N\!-\!1]$} }
\psfrag{d5}{\hspace{-5mm}\scriptsize{$d_{M\!-\!1}[N\!-\!1]$} }
\psfrag{e0}{\hspace{-0.3  mm}\scriptsize{$\omega_0^0$} }
\psfrag{e1}{\hspace{-0.7 mm}{\scriptsize$\omega_0^{\hspace{-0.3mm}M\!-\!1}$} }
\psfrag{e2}{\hspace{-0.3  mm}\scriptsize{$\omega_{1}^0$} }
\psfrag{e3}{\hspace{-0.7  mm}{\scriptsize$\omega_{1}^{M-1}$} }
\psfrag{e4}{\hspace{-0.7 mm}\scriptsize{$\omega_{\hspace{-0.3mm}N\!-\!1}^0$} }
\psfrag{e5}{\hspace{-1.2 mm}{\scriptsize$\omega_{\hspace{-0.3mm}N\!-\!1}^{\hspace{-0.3mm}M\!-\!1}$} }
\psfrag{p/s}{\footnotesize{P/S} }
\psfrag{and}{\footnotesize{\&} }
\psfrag{CP}{\hspace{-0.4  mm}\footnotesize{CP} }
\psfrag{Add}{\hspace{-0.6mm}\footnotesize{Addition} }
\centering
\includegraphics[scale=0.26]{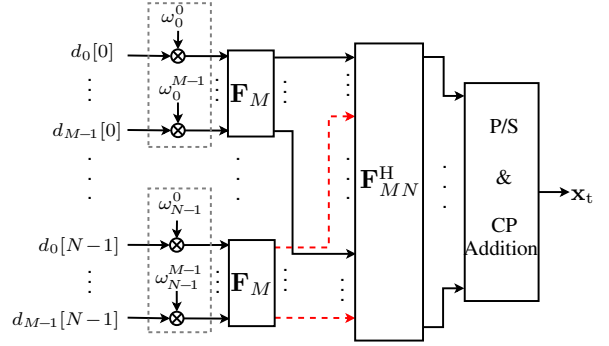} 
\vspace{-0.3 cm}
\caption{SC-FDMA-based delay-Doppler modulator.}
\vspace{-0.4 cm}
\label{fig:OTFS_Tx1}
\end{figure}

\begin{figure}
\psfrag{F_MH}{ \hspace{-1.3mm}$\mathbf{F}_M^{\rm H}$}
\psfrag{F_MN}{ \hspace{-2.7 mm}$\mathbf{F}_{MN}$}
\psfrag{M}{{$M$} }
\psfrag{r}{\hspace{-0.8 mm}{\small$\bm{r}_{\rm{t}}$} }
\psfrag{d0}{\hspace{-0.5  mm}\scriptsize{$\widetilde{d}_0[0]$} }
\psfrag{d1}{\hspace{-0.5  mm}\scriptsize{$\widetilde{d}_{M-1}[0]$} }
\psfrag{d2}{\hspace{-0.5  mm}\scriptsize{$\widetilde{d}_0[1]$} }
\psfrag{d3}{\hspace{-0.5  mm}\scriptsize{$\widetilde{d}_{M-1}[1]$} }
\psfrag{d4}{\hspace{-0.5  mm}\scriptsize{$\widetilde{d}_0[N-1]$} }
\psfrag{d5}{\hspace{-0.5  mm}\scriptsize{$\widetilde{d}_{M-1}[N-1]$} }
\psfrag{e0}{\hspace{-0.3  mm}\tiny{$\large(\hspace{-0.2  mm}\omega_{0}^{0}\hspace{-0.2  mm}\large)^*$} }
\psfrag{e1}{\hspace{-0.4 mm}{\tiny$\large(\hspace{-0.4  mm}\omega_{0}^{\!M-1}\hspace{-0.4  mm}\large)^*$} }
\psfrag{e2}{\hspace{-0.3  mm}\tiny{$\large(\hspace{-0.2  mm}\omega_{1}^{0}\hspace{-0.2  mm}\large)^*$} }
\psfrag{e3}{\hspace{-0.7 mm}{\tiny$\large(\hspace{-0.2  mm}\omega_{1}^{\!M-1}\!\large)^*$} }
\psfrag{e4}{\hspace{-0.7 mm}\tiny{$\large(\hspace{-0.2  mm}\omega_{\!N-1}^{0}\hspace{-0.2  mm}\large)^*$} }
\psfrag{e5}{\hspace{-0.7 mm}{\tiny$\large(\hspace{-0.4  mm}\omega_{\!N-1}^{\!M-1}\hspace{-0.4  mm}\large)^*$} }
\psfrag{p/s}{\hspace{-0.4  mm}\footnotesize{S/P} }
\psfrag{and}{\hspace{0.4  mm}\footnotesize{\&} }
\psfrag{CP}{\hspace{-1.3  mm}\footnotesize{CP} }
\psfrag{Rmv}{\hspace{-1.5  mm}\footnotesize{Remove} }
\centering
{\hspace{-9 mm}\includegraphics[scale=0.25]{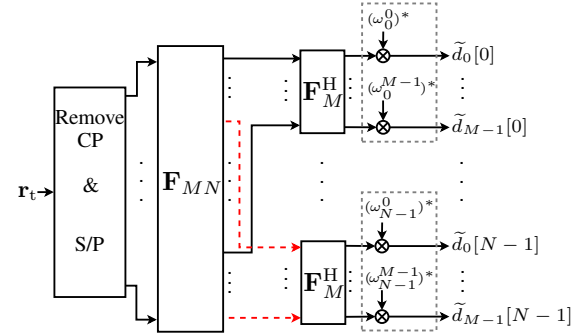} }
\vspace{-0.15 cm}
\caption{SC-FDMA-based delay-Doppler demodulator.}
\vspace{-0.7 cm}
\label{fig:OTFS_Rx1}
\end{figure}

In OTFS modulation, two approaches are commonly taken to form the transmit signal; (i) the delay-Doppler domain data symbols are transformed to the time-frequency domain and then the delay-time domain signal is obtained after applying Heisenberg transform \cite{b2}. (ii) the delay-Doppler domain data symbols are directly converted to delay-time domain using Zak transform \cite{b3, b14, b19}. 
The recently introduced approach in  \cite{b9} and \cite{b15} proposes an alternative transformation pathway, i.e., converting the delay-Doppler domain symbols into the frequency-Doppler domain before transforming them into the delay-time domain \cite{b9}, see the path shown by red arrows in Fig.~\ref{fig:transform}.  
The modulator and demodulator structures for this alternative approach are shown in Figs.~\ref{fig:OTFS_Tx1} and \ref{fig:OTFS_Rx1}, respectively, with the following mathematical representation. 

We consider quadrature amplitude modulated (QAM) data symbols drawn from a zero-mean, independent, and identically distributed process with unit variance. These symbols are arranged on a grid in the delay-Doppler domain with delay and Doppler spacings of $\Delta\tau$ and $\Delta\nu$, respectively. Each OTFS frame consists of $M$ delay bins and $N$ Doppler bins.
Thus, each delay block has the duration of $T = M\Delta\tau$ and the Doppler spacing is defined as $\Delta\nu=\frac{1}{NT}=\frac{1}{MN\Delta\tau}$. Additionally, $\mathbf{D}\in \mathbb{C}^{M\times N}$ represents the data matrix with the elements $d_m[n]$ for  $m = 0, . . . , M - 1$ and $n = 0, . . . , N - 1$ where $m$ and $n$ denote the delay and Doppler indices, respectively. 

Considering the second approach, the ZAK transform is applied to the delay-Doppler domain signal, converting it into the delay-time domain, as 
\begin{equation} \label{eqn:st1}
\mathbf{s}_{\rm t} = (\mathbf{F}_{N}^{\rm H}\otimes \mathbf{I}_{M})\mathbf{d},
\end{equation}
where $\bm{d}$ represents the vectorized form of the matrix $\bm{D}$.
\begin{figure*}[!t]
\psfrag{F_M}{ \hspace{-1.4mm}$\mathrm{F}_M$}
\psfrag{F_MN}{ \hspace{-1.7 mm}$\mathrm{F}_{MN}^{\rm H}$}
\psfrag{M}{{$M$} }
\psfrag{x}{{\small$\bm x$} }
\psfrag{d0}{\hspace{0.0mm}\scriptsize{$d_0[0]$} }
\psfrag{d1}{\hspace{-1.75mm}\scriptsize{$d_{M\!-\!1}[0]$} }
\psfrag{d2}{\hspace{0.0mm}\scriptsize{$d_0[1]$} }
\psfrag{d3}{\hspace{-1.9mm}\scriptsize{$d_{M\!-\!1}[1]$} }
\psfrag{d4}{\hspace{-4.0mm}\scriptsize{$d_0[N\!-\!1]$} }
\psfrag{d5}{\hspace{-6.8mm}\scriptsize{$d_{M\!-\!1}[N\!-\!1]$} }
\psfrag{e0}{\hspace{-0.3  mm}\tiny{$\omega_0^0$} }
\psfrag{e1}{\hspace{-0.7 mm}{\tiny$\omega_0^{\hspace{-0.3mm}M\!-\!1}$} }
\psfrag{e2}{\hspace{-0.3  mm}\tiny{$\omega_1^0$} }
\psfrag{e3}{\hspace{-0.7 mm}{\tiny$\omega_1^{M-1}$} }
\psfrag{e4}{\hspace{-0.7 mm}\tiny{$\omega_{N\!-\!1}^0$} }
\psfrag{e5}{\hspace{-1.2 mm}{\tiny$\omega_{N\!-\!1}^{\hspace{-0.3mm}M\!-\!1}$} }
\psfrag{p/s}{\hspace{-0.4mm}\footnotesize{P/S} }
\psfrag{UFMC Modulator}{\hspace{-0.2mm}\footnotesize{UFMC Modulator} }
\psfrag{V}{\footnotesize{$\bm{V}^{\rm{H}}$} }
\psfrag{W}{\footnotesize{$\bm{W}$} }
\psfrag{P}{\footnotesize{$\bm{P}$} }
\psfrag{s}{\footnotesize{$\bm{s}^{\mathtt{GF}}_{\rm{t}}$} }
\psfrag{IFFT}{\footnotesize{IDFT} }
\psfrag{Filtering}{\footnotesize{Filtering} }
\psfrag{Predistortion}{\hspace{-0.6mm}\footnotesize{Predistortion} }
\centering
\includegraphics[scale=0.25]{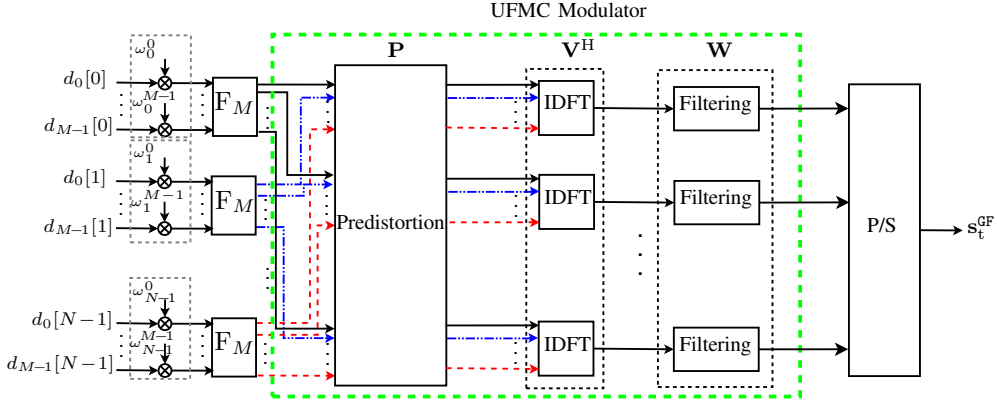}
\caption{The proposed GF-OTFS modulator.}
\vspace{-0.2 cm}
\label{fig:scfdma_Tx}
\end{figure*}

According to \cite{b9}, OTFS modulation can equivalently be implemented using SC-FDMA by applying a twiddling factor to the transmit and receive data symbols.
To this end, we implement (\ref{eqn:st1}) using SC-FDMA.
By leveraging the Cooley-Tukey general factorization, a large DFT can be decomposed into two smaller DFTs as
\begin{equation} \label{eqn:ct}
\bm{F}_{MN} = \bs{\Psi} ( \bm{I}_{N} \otimes \bm{F}_{M}) \bs{\Omega} (\bm{F}_{N}\otimes \bm{I}_{M}),
\end{equation}
where $\bs{\Omega}\!\!=\!\!{\rm{diag}} \{\bs{\omega}_0,\ldots,\bs{\omega}_{N-1}\}$ for $\bs{\omega}_n\!\!=\!\![\omega_n^0,\ldots,\omega_n^{M-1}]^{\rm T}$ and $\omega_n^m=e^{-j\frac{2 \pi mn}{MN}}$.
Furthermore, $\bs{\Psi}=[\bs{\Psi}_0,\ldots,\bs{\Psi}_{M-1}]^{\rm{T}}$ is an interleaving matrix in which $\bs{\Psi}_m= {\rm{CircShift}} \big( (\bm{I}_N \otimes \bs{\psi}),m \big)$ for $m\!=\!0,\ldots,M\!-\!1$ and $\bs{\psi}=[1,\bm{0}^{\rm{T}}_{M-1}]^{\rm{T}}$.
Assume $\bs{\Gamma} = \bs{\Psi} (\bm{I}_{N} \otimes \bm{F}_{M}) \bs{\Omega}$.
The matrix $\bs{\Gamma}$ is unitary because $\bs{\Psi}$, $(\bm{I}_{N} \otimes \bm{F}_{M})$, and $\bs{\Omega}$ are all unitary matrices, and which implies $\mathbf\Gamma\mathbf\Gamma^{\rm H} \!=\! \mathbf{I}_{MN}$.
Using (\ref{eqn:ct}), equation (\ref{eqn:st1}) can be rewritten as
\begin{align} \label{eqn:st}
\mathbf{s}_{\rm t} &=(\mathbf{F}_{N}^{\rm H}\otimes \mathbf{I}_{M}) \mathbf\Gamma^{\rm H}\mathbf\Gamma\mathbf{d} \nonumber \\
&=(\mathbf{F}_{N}^{\rm H}\otimes \mathbf{I}_{M}) \Big( \bs{\Psi} (\mathbf{I}_{N} \otimes \mathbf{F}_{M}) \boldsymbol{\Omega} \Big)^{\rm H} \bs{\Psi} (\mathbf{I}_{N} \otimes \mathbf{F}_{M}) \boldsymbol{\Omega} \mathbf{d} \nonumber \\ &= \mathbf{F}_{MN}^{\rm H} \bs{\Psi} ( \mathbf{I}_{N} \otimes \mathbf{F}_{M}) \boldsymbol{\Omega} \mathbf{d} = \mathbf{F}_{MN}^{\rm H}  \bm{s}_{\rm{f}},
\end{align}
where ${\bf{s}_{\rm{f}}} = \bs{\Psi} (\bm{I}_N \otimes \bm{F}_M) \bs{\Omega} \bm{d}$.  
As shown in Fig.~\ref{fig:OTFS_Tx1}, one may realize that the third term in (\ref{eqn:st}) can be implemented by transmitting the twiddled data in the delay-Doppler domain, i.e., $\bs{\Omega} \bm{d}$ through an SC-FDMA modulator \cite{b15}.
Then, a CP of length ${L_{\rm cp}} \geq L-1$ is added at the start of each OTFS frame to ensure inter-block interference-free communication, where $L$ is the channel length.
Thus, the transmit signal can be represented as
\begin{equation}
 \mathbf{x}_{\rm{t}} = \mathbf{A}_{\rm cp} \mathbf{s}_{\rm t},
\end{equation}
where $\mathbf{A}_{\rm cp} = [{\mathbf G}_{\rm cp}^{\rm T}, {\mathbf{I}_{MN}^{\rm T}}]^{\rm T} $ is the CP addition matrix where the ${L}_{\rm cp} \times MN$ matrix $\mathbf{G}_{\rm cp}$ is formed by taking the last ${L}_{\rm cp}$  rows of the identity matrix $\mathbf{I}_{MN}$.
The signal then propagates through a linear time-varying (LTV) channel. After CP removal, the received signal is given by
\begin{equation}
\mathbf{r}_{\rm{t}} = \mathbf{H} \mathbf{x}_{\rm{t}} + \boldsymbol{\eta},
 \end{equation}
where $\bm{H}$ is the channel matrix in the delay-time domain with the elements $[\bm{H}]_{i,j}=h[i-j,i]$ for $i,j=0,\ldots,K-1$ and $K=MN+L_{\rm{cp}}$.
Furthermore, $h[j,i]$ is the channel gain at the delay tap $j$ and sample $i$, and
$\boldsymbol{\eta} \sim \mathcal{CN}(0, \sigma_{\eta}^2 \mathbf{I}_{K})$ is the additive white Gaussian noise (AWGN) with the variance $\sigma_{\eta}^2$.

At the receiver, as shown in Fig.~\ref{fig:OTFS_Rx1}, the signal is processed through an SC-FDMA demodulator.
Specifically, the CP is first removed by multiplying it with the CP removal matrix, $\mathbf{B}_{\rm cp} = [\mathbf{0}_{MN \times {L_{\rm cp}}}, \mathbf{I}_{MN}]$.
The signal then passes through the OFDM demodulator, followed by symbol recovery of dimension ($M$, $N$) via de-interleaving. 
Consequently, an IDFT is applied along the frequency dimension to transform the signal from the frequency-Doppler domain to the delay-Doppler domain.
Finally, the twiddled signal is rotated back to its original phase by multiplying with $\bs{\Omega}^{\rm{H}}$.
This demodulator process can be represented as
\begin{align} \label{eqn:H_dd}
\widetilde{\bm{d}} = \bs{\Omega}^{\rm{H}}(\mathbf{I}_{N} \otimes \mathbf{F}^{\rm{H}}_{M})  \bs{\Psi}^{\rm H} \mathbf{F}_{MN} \mathbf{B}_{\rm cp} \mathbf{r}_{\rm{t}} = \bm{H}_{\rm{DD}} \bm{d} + \bs{\eta}_{\rm{DD}},
\end{align}
where the equivalent channel matrix in the delay-Doppler domain is represented as $\mathbf{H}_{\rm DD}  = \bs{\Gamma}^{\rm{H}} \bm{F}_{MN} \bm{H}_{\rm{DT}} \bm{F}^{\rm{H}}_{MN} \bs{\Gamma} \!=\! \bs{\Omega}^{\rm{H}} (\mathbf{I}_{N} \!\otimes\! \mathbf{F}_{\!M}^{\rm H}) \bs{\Psi}^{\rm H} \mathbf{F}_{\!\!MN} \mathbf{H}_{\rm DT} \mathbf{F}_{\!\!MN}^{\rm H} \bs{\Psi} (\mathbf{I}_{N} \!\otimes\! \mathbf{F}_{\!\!M} )
\bs{\Omega}$, and $\bm{H}_{\rm{DT}}=\mathbf{B}_{\rm cp} \mathbf{H} \mathbf{A}_{\rm cp}$.
Additionally, $\bs{\eta}_{\rm{DD}}$ is the AWGN in the delay-Doppler domain that can be obtained by $\bs{\Gamma}^{\rm{H}} \mathbf{F}_{MN} \mathbf{B}_{\rm cp} \bs{\eta}$.

As described in this Section, the OTFS implementation using SC-FDMA utilizes an OFDM modulator to transform the signal between the frequency-Doppler and delay-time domains.
However, the OFDM modulator struggles to handle the Doppler effect in doubly selective channels effectively.
To address this limitation, the OFDM modem can be replaced with a more robust alternative, i.e., UFMC that mitigates the interference between the frequency-Doppler bins.

\section{Proposed Globally Filtered OTFS}

The matrices $\bm{F}_{MN}^{\rm{H}}$ and $\bm{F}_{MN}$ in (\ref{eqn:H_dd}) constitute an OFDM modem in the SC-FDMA-based OTFS for frequency-Doppler transmission.
This modem transforms the data symbols between frequency-Doppler and delay-time domains.
Due to the interleaving block in this system, consecutive samples in the frequency-Doppler domain prior to modulation correspond to frequency-Doppler bins.
This ability to control the Doppler dimension through the modulator provides a valuable opportunity to employ a more effective modem to address issues caused by IDI.
Thus, the OFDM modem for the frequency-Doppler transmission in this system can be substituted with alternative modems, such as UFMC, to mitigate interferences caused by both delay and Doppler effects, thereby enhancing the overall system performance.

A UFMC modulator groups $N_{\rm sc} \!\!=\!\! N_{\rm rb} \times N_{\rm sc}^{\rm rb}$ subcarriers into  
$N_{\rm rb}$ subbands, each including $N_{\rm sc}^{\rm rb}$ subcarriers.
The filtering process can be achieved by performing an IDFT on the signal of each subband to transform it into the time domain, followed by the filtering.
The smooth edges of the filter used for each subband in time, significantly suppress OOB emissions in the frequency domain. Additionally, this filtering mitigates interference from neighboring subcarriers in two adjacent subbands.
Finally, all subband signals are summed to form the transmit time-domain signal \cite{b11}.
Furthermore, before the subbands are formed, a predistortion stage may be applied to counteract the effects of filtering, specifically addressing the non-linearities introduced by subband filtering \cite{b16}.

In this Section, we propose a multiplexing technique that employs both SC-FDMA-based OTFS and UFMC. For reasons that will become clear shortly, we name this technique GF-OTFS.
In this technique, the interleaving stage of OTFS implemented with SC-FDMA arranges the frequency-Doppler bins in adjacency.
Following this, the UFMC modulator groups these adjacent bins into subbands and applies filtering.
This grouping and filtering, which is applied to adjacent frequency-Doppler bins, mitigates the IDI and intercarrier interference. 
Essentially, this filtering is applied across the entire OTFS frame on the frequency-Doppler domain, that is the reason behind the naming the proposed technique as GF-OTFS.
As a result, the interference is reduced in both frequency (delay) and Doppler dimensions.
In contrast, in \cite{b7}, UFMC is applied along the delay dimension to suppress delay interference, which is ineffective for reducing the Doppler effect on the signal.

As shown in Fig.~\ref{fig:scfdma_Tx}, to apply the UFMC modulator to the signal in (\ref{eqn:st}), the OFDM modulator should be replaced with the UFMC modulator.
In the case of using the UFMC modulator without predistortion, the modulated signal can be obtained by
\begin{equation} \label{eqn:u_mod0}
\bm{s}_{\rm{t,0}} = \bm{T}_0 \bm{s}_{\rm{f,0}},   
\end{equation}
where $N_{\rm{sc}}=MN$ and $\mathbf{T}_0 = \bm{WV^{\rm{H}}}$. 
In $\bm{T}_0$, the matrix $\mathbf{V} = {\rm diag}\{\mathbf{V}_0, \ldots, \mathbf{V}_{N_{\rm{rb}}-1}\}$ consists submatrices $\mathbf{V}_{i} \in \mathbb{C}^{N_{\rm sc} \times N_{\rm sc}^{\rm rb}}$.
Each $\mathbf{V}_{i}$ is constructed from the $N_{\rm{sc}}^{\rm{rb}}$ columns of $\bm{F}_{N_{\rm{sc}}}$, starting from the $(i N^{\rm{rb}}_{\rm{sc}})^{\rm{th}}$ column, for $i=0,\ldots,N_{\rm{rb}}-1$.
Furthermore, $\mathbf{W} = [\mathbf{W}_0,\ldots, \mathbf{W}_{N_{\rm rb}-1}]$ is filter matrix covering all subbands, where
$\mathbf{W}_i = \mathcal{T}_{N_{\rm sc}+L^{\mathtt{GF}}_{\rm{f}}-1\times N_{\rm sc}} \{ \bs{\Phi}_i \bm{w} \}$. 
The vector $\bm{w}$ is a finite
impulse response (FIR) filter of length $L^{\mathtt{GF}}_{\rm{f}}$, and
$\bs{\Phi}_i={\rm{diag}}[e^{j \frac{2 \pi \alpha_i}{N_{\rm{sc}}}0},\ldots, e^{j \frac{2 \pi \alpha_i}{N_{\rm{sc}}} (L^{\mathtt{GF}}_{\rm{f}}-1)}]$ applies a circular frequency shift to the FIR filter by $\alpha_i=(i+0.5)N_{\rm sc}^{\rm rb}-0.5$, where $i=0,\ldots,N_{\rm{rb}}-1$ represents the subband index.

\begin{figure}[!t]
\psfrag{F_M}{ \hspace{-1.3mm}$\mathrm{F}_M^{\rm H}$}
\psfrag{M}{{$M$} }
\psfrag{r}{{\small$\bm r$} }
\psfrag{d0}{\hspace{-0.0 mm}\scriptsize{$\widetilde{d}^{\mathtt{\,GF}}_0[0]$} }
\psfrag{d1}{\hspace{-0.0  mm}\scriptsize{$\widetilde{d}^{\mathtt{\,GF}}_{M\!-\!1}[0]$} }
\psfrag{d2}{\hspace{-0.0  mm}\scriptsize{$\widetilde{d}^{\mathtt{\,GF}}_0[1]$} }
\psfrag{d3}{\hspace{-0.0  mm}\scriptsize{$\widetilde{d}^{\mathtt{\,GF}}_{M\!-\!1}[1]$} }
\psfrag{d4}{\hspace{-0.0  mm}\scriptsize{$\widetilde{d}^{\mathtt{\,GF}}_0[N\!-\!1]$} }
\psfrag{d5}{\hspace{-0.0  mm}\scriptsize{$\widetilde{d}^{\mathtt{\,GF}}_{M\!-\!1}[N\!-\!1]$} }
\psfrag{e0}{\hspace{-0.3 mm}\tiny{$(\hspace{-0.2  mm}\omega_{0}^{0}\hspace{-0.2  mm}\large)^*$} }
\psfrag{e1}{\vspace{0.2mm} \hspace{-1.9 mm}{\tiny$(\hspace{-0.5  mm}\omega_{0}^{\!M\!-\!1}\hspace{-0.4 mm})^*$} }
\psfrag{e2}{\hspace{-0.3 mm}\tiny{$(\hspace{-0.2  mm}\omega_{1}^{0}\hspace{-0.2 mm})^*$} }
\psfrag{e3}{\hspace{-0.7 mm}{\tiny$(\hspace{-0.2  mm}\omega_{1}^{\!M-1}\!)^*$} }
\psfrag{e4}{\hspace{-0.7 mm}\tiny{$(\hspace{-0.2 mm}\omega_{\!N-1}^{0}\hspace{-0.2 mm})^*$} }
\psfrag{e5}{\hspace{-0.7 mm}{\tiny$(\hspace{-0.4 mm}\omega_{\!N-1}^{\!M-1}\hspace{-0.4 mm})^*$} }
\psfrag{p/s}{\hspace{-0.0  mm}\footnotesize{S/P} }
\psfrag{and}{\hspace{0.4  mm}\footnotesize{\&} }
\psfrag{CP}{\hspace{-1.3  mm}\footnotesize{CP} }
\psfrag{Rmv}{\hspace{-1.2  mm}\footnotesize{Remove} }
\psfrag{UFMC Demodulator}{\hspace{-1.1mm}\footnotesize{UFMC Demodulator} }
\psfrag{Os}{$\bm{F}_{2N_{\rm sc}}$}
\psfrag{Od}{{\small$\bm{E}$} }
\psfrag{FFT}{\hspace{-0.8  mm}\footnotesize{DFT} }
\psfrag{Carrier}{\vspace{1.5mm} \footnotesize{Carrier} }
\psfrag{Selection}{\vspace{0.5mm}\footnotesize{Selection} }
\psfrag{z}{\footnotesize{Zero} }
\psfrag{x}{\footnotesize{X} }
\psfrag{p}{\footnotesize{padding} }
\psfrag{r}{\footnotesize{\!\!$\bm{r}^{\mathtt{GF}}_{\rm{t}}$} }
\psfrag{Demodulator}{\footnotesize{Demodulator} }

\includegraphics[scale=0.227]{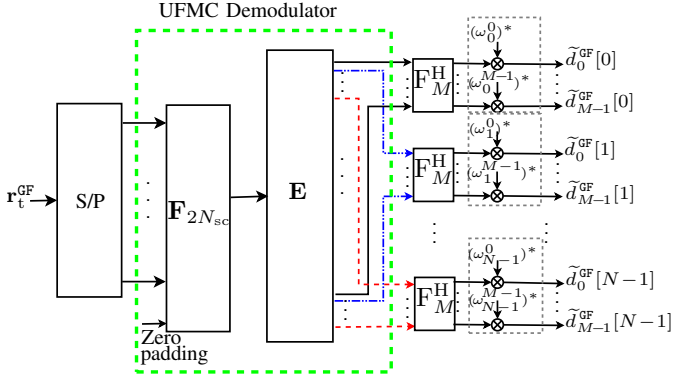}

\vspace{-0.3 cm}
\caption{The proposed GF-OTFS demodulator.}
\vspace{-0.8 cm}
\label{fig:scfdma_Rx1}
\end{figure}

To ensure the transmitted symbols remain as close as possible to their intended form in the frequency domain, a predistortion stage is required to be applied to the signal before the UFMC filtering process.
To analyze the distortive effect of a UFMC modem, we modulate a signal consisting of all ones in the frequency-Doppler domain using the UFMC modulator presented in (\ref{eqn:u_mod0}). Specifically, let $\bm{s}_{\rm{f,0}}=\bs{1}_{N_{\rm{sc}}}$, resulting in $\widetilde{\bm{s}}_{\rm{t,0}}= \bm{T}_0 \bs{1}_{N_{\rm{sc}}}$.
Subsequently, the UFMC demodulator is applied to revert the modulated signal back to its original state.
This process allows us to account for all stages that introduce distortions to the original signal, which can collectively be considered as the UFMC modem's effect. By understanding these effects, we can introduce an additional stage of predistortion at the transmitter to effectively cancel out the impact of UFMC modulation.

As the initial step in canceling out the distortive effect of the UFMC modem, the modulation matrix $\mathbf{T}_0$ is normalized by the power of $\widetilde{\bm{s}}_{\rm{t,0}}$, expressed as
\begin{equation} \label{eqn:norm}
\mathbf{T}_{\rm n}
= \frac{\mathbf{T}_0}{\Big(\frac{\widetilde{\bm{s}}_{\rm{t,0}}^{\rm{H}} \widetilde{\bm{s}}_{\rm{t,0}}}{N_{\rm sc}+L^{\mathtt{GF}}_{\rm f}-1} \Big)^{0.5}}.
\end{equation}
In the UFMC demodulator, the signal is first extended by zero-padding using $\mathbf{Z} = [\mathbf{I}^{\rm{T}}_{N_{\rm sc}+L^{\mathtt{GF}}_{\rm f}-1},\mathbf{0}^{\rm{T}}_{(N_{\rm sc}-L^{\mathtt{GF}}_{\rm f}+1) \times (N_{\rm sc}+L^{\mathtt{GF}}_{\rm f}-1)}]^{\rm{T}}$,
that prepares it for the application of an oversampled DFT, i.e., $\mathbf{F}_{2N_{\rm sc}}$.
Then,  the downsampling matrix $\mathbf{E} \in \mathbb{C}^{N_{\rm sc} \times 2N_{\rm sc}}$, consisting of every other row of $\mathbf{I}_{2N_{\rm sc}}$, is applied to the signal.
These stages of the UFMC demodulator are represented by $\bm{R}_{\rm{u}}=\mathbf{E}\mathbf{F}_{2N_{\rm sc}} \mathbf{Z}$ and are applied to the modulated signal as
\begin{equation} \label{eqn:u_demod0}
\widetilde{\bm{s}}_{\rm{f,0}} = \bm{R}_{\rm{u}} \widetilde{\bm{s}}_{\rm{t,0}} = \bm{R}_{\rm{u}} \bm{T}_0 \bs{1}_{N_{\rm{sc}}},
\end{equation}
which is the original signal $\bs{1}_{N_{\rm{sc}}}$ being affected by the UFMC modem.
Thus, the predistortion matrix can be obtained by dividing the transmit signal $\bs{1}_{N_{\rm{sc}}}$ by the signal processed through the UFMC modem in (\ref{eqn:u_demod0}).
The predistortion matrix in the frequency-Doppler domain is expressed as
\begin{equation} \label{eqn:dist}
\mathbf{P} = \mathrm{diag}\Big\{ \mathbf{1}_{N_{\rm sc}}\oslash {\frac{\widetilde{\bm{s}}_{\rm{f,0}}}{\frac{1}{N_{\rm sc}}\sum\limits_{k=0}^{N_{\rm sc}-1}|\widetilde{\bm{s}}_{\rm{f,0}}[k]|}} \Big\}.
\end{equation}
Considering (\ref{eqn:norm}) and (\ref{eqn:dist}), the transmit signal after passing through a UFMC modulator with predistortion can be obtained from (\ref{eqn:u_mod0}), where the UFMC modulator $\bm{T}_0$ is replaced by $\bm{T}_{\rm{u}}=\bm{T}_{\rm{n}} \bm{P}$.

As shown in Fig.~\ref{fig:scfdma_Tx}, the transmit signal for GF-OTFS, incorporating the predistortion stage for the UFMC component of the proposed modulator, can be expressed as
\begin{equation} \label{eqn:u_mod}
\bm{s}^{\mathtt{GF}}_{\rm{t}} = \bm{T}_{\rm{u}} \bm{s}_{\rm{f}} = \bm{T}_{\rm{u}} \bs{\Psi} ( \mathbf{I}_{N} \otimes \mathbf{F}_{M}) \boldsymbol{\Omega} \mathbf{d}, 
\end{equation}
where $\bm{s}_{\rm{f}} = \bs{\Psi} (\bm{I}_N \otimes \bm{F}_M) \bs{\Omega} \bm{d}$ represents the signal in the frequency-Doppler domain, derived using the SC-FDMA modulator in (\ref{eqn:st}).
The UFMC transmit signal does not need a CP \cite{b11}, and (\ref{eqn:u_mod}) is transmitted through the LTV channel.
The received signal can be expressed as 
\begin{equation}
\mathbf{r_{\rm{t}}^{\mathtt{GF}}} = \overline{\mathbf{H}} \mathbf{s}_{\rm t}^{\mathtt{GF}} + \overline{\boldsymbol{\eta}},
 \end{equation}
where $\overline{\bm{H}}$ is the channel matrix, representing the LTV channel in the delay-time domain with the elements $[\overline{\bm{H}}]_{i,j}=h[i-j,i]$ for $i,j=0,\ldots,\overline{K}-1$ and $\overline{K}=N_{\rm{sc}}+L^{\mathtt{GF}}_{\rm{f}}$. Additionally,
$\overline{\bs{\eta}} \sim \mathcal{CN}(0, \sigma_{\eta}^2 \mathbf{I}_{\overline{K}})$ is the AWGN.

\renewcommand{\arraystretch}{1.4}
\begin{table}
\vspace{-0.2cm}
\centering
\caption{Simulation Parameters}\vspace{-0.2cm}
\resizebox{0.489\textwidth}{!}
{\begin{tabular}{| c || c |}
\hline\hline
\textbf{Simulation Parameter}
& \textbf{Value}
\\ \hline \hline
OTFS frame
& $M=64$ and $N=8$
\\ \hline
Carrier frequency
& $f_c=5.9~\rm{GHz}$
\\ \hline
Bandwidth
& $B=1.92~\rm{MHz} \, and \, 10~\rm{MHz}$
\\ \hline
Channel
& Tapped delay line clustered (TDL-C) model \cite{b17} and $L\!=\!5$
\\ \hline
Speed
& $v=500~\rm{km/h}$
\\ \hline
UFMC Frame
& $N_{\rm{sc}}\!=\!MN$ and $N_{\rm{sc}}^{\rm{rb}}\!=\!4$ for both GF-OTFS and DR-UFMC
\\ \hline
FIR filter
& Chebyshev filter
\\ \hline
Length of filter
&  $L_{\rm f}^{\rm GF} \!=\! \frac{MN}{4}\!+\!1$ for GF-OTFS and $L_{\rm f}^{\rm DU}\!=\!20$ for DR-UFMC

\\ \hline
CP length for RW-OTFS
& $\frac{MN}{4}$
\\ \hline\hline
\end{tabular}
\label{tab1}
}
\vspace{-0.3cm}
\end{table}

At the receiver, the demodulation process for the proposed GF-OTFS is essentially the inverse of the transmission procedure as shown in Figure \ref{fig:scfdma_Rx1}. The received signal first passes through the UFMC demodulator, followed by de-interleaving to recover the symbols of dimension $(M,N)$.
Next, IDFT is applied along the frequency dimension to transform the signal from the frequency-Doppler domain to the delay-Doppler domain.
The received data symbols in the delay-Doppler domain are then untwiddled by multiplying them by $\bs{\Omega}^{\rm{H}}$.
This proposed GF-OTFS demodulator can be formulated as
\begin{align}
\widetilde{\mathbf{d}}^{\mathtt{GF}} = \bs{\Omega}^{\rm{H}} (\mathbf{I}_{N} \otimes\mathbf{F}_{M}^{\rm H}) \bs{\Psi}^{\rm H} \mathbf{R}_{\rm u} \mathbf{r_{\rm{t}}^{\mathtt{GF}}} = \bm{H}_{\rm{DD}}^{\mathtt{GF}} \bm{d} + \overline{\bs{\eta}}_{\rm{DD}},
\end{align}
recalling that $\bm{R}_{\rm{u}}=\mathbf{E}\mathbf{F}_{2N_{\rm sc}}\mathbf{Z}$ is the UFMC demodulation matrix.
The delay-Doppler domain equivalent channel matrix for GF-OTFS can be represented as $\mathbf{H}_{\rm DD}^{\mathtt{GF}} = \bs{\Gamma}^{\rm{H}} \bm{H}^{\mathtt{GF}}_{\rm{FD}} \bs{\Gamma} = \bs{\Omega}^{\rm{H}}  (\mathbf{I}_{N} \otimes \mathbf{F}_{M}^{\rm H}) \bs{\Psi}^{\rm H} \mathbf{H}^{\mathtt{GF}}_{\rm FD} \bs{\Psi} (\mathbf{I}_{N} \otimes \mathbf{F}_{M}) \bs{\Omega}$.
Here, $\bm{H}^{\mathtt{GF}}_{\rm{FD}}=\mathbf{R}_{\rm u} \overline{\mathbf{H}} \mathbf{T}_{\rm u}$ represents the GF-OTFS equivalent channel in the frequency-Doppler domain, and the AWGN in the delay-Doppler domain is given by $\overline{\bs{\eta}}_{\rm{DD}}= \bs{\Gamma}^{\rm{H}} \bm{R}_{\rm{u}} \overline{\bs{\eta}}$.
Finally, the symbol detection can also be performed using the minimum mean square error (MMSE) technique \cite{b18}, as follows
\begin{equation}
\mathbf{\widehat{d}^{\mathtt{GF}}} = \Big( (\mathbf{H}^{\mathtt{GF}}_{\rm DD})^{\rm H} \mathbf{H}^{\mathtt{GF}}_{\rm DD} + \sigma_{\eta}^2 \mathbf{I}_{MN} \Big)^{-1} (\mathbf{H}^{\mathtt{GF}}_{\rm DD})^{\rm H} {\widetilde{\mathbf{d}}^{\mathtt{GF}}}.
\end{equation}

\section{Numerical Results}
\label{sec:results}

In this section, we evaluate the performance of the proposed GF-OTFS by comparing it with other multiplexing techniques through a series of numerical simulations.
The simulation parameters for each technique are provided in Table \ref{tab1}.

\begin{figure}[t!]
\centering
\includegraphics[scale=0.31]{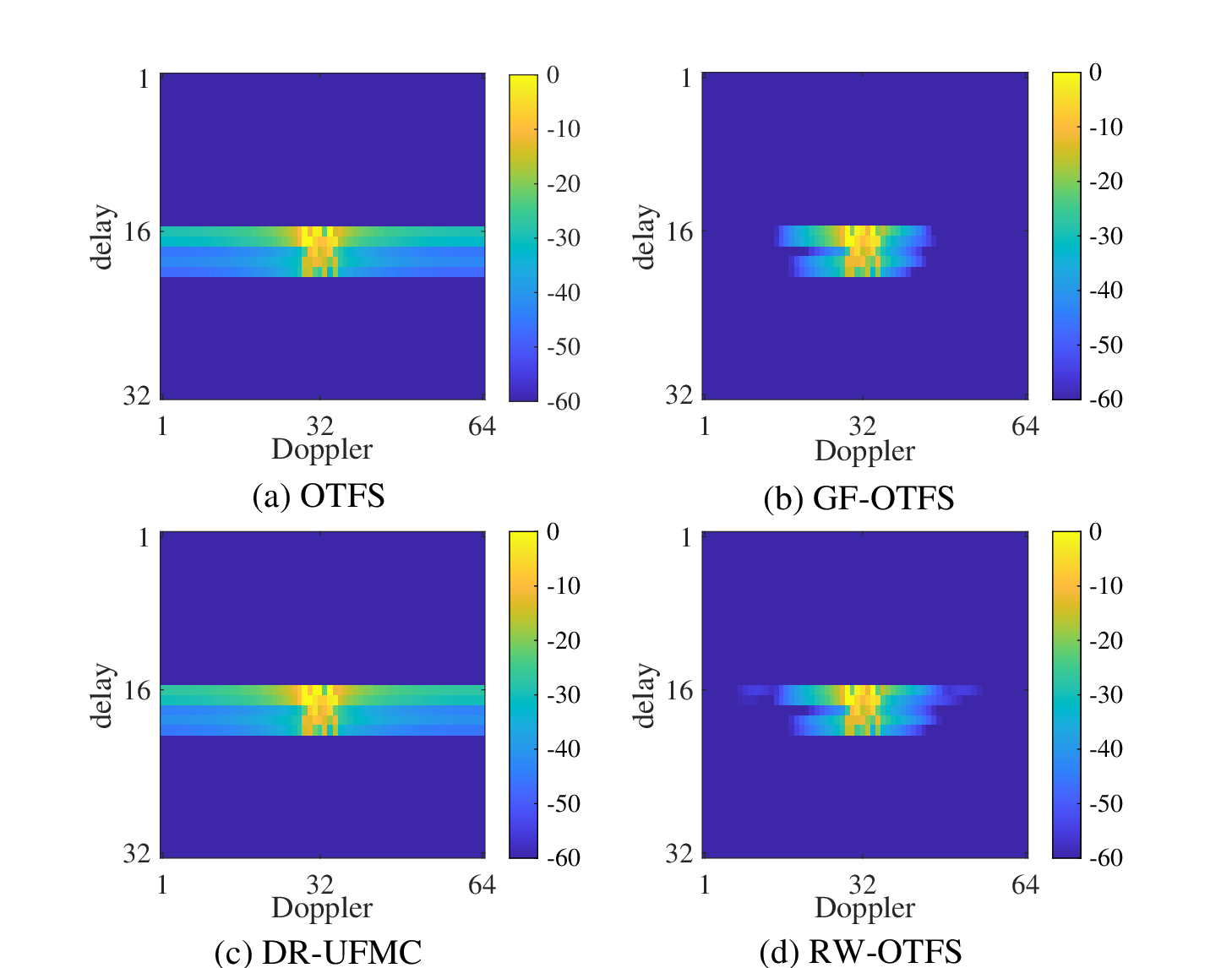} 
\vspace{-0.3cm}
\caption{Impulsive samples received in the delay-Doppler domain under fractional Doppler shifts.}
\label{fig:spread}
\vspace{-0.35cm}
\end{figure}

We first activate only the center delay-Doppler bin as the transmit symbol and pass the signal through the channel. Both the proposed GF-OTFS and RW-OTFS, as shown in Fig.~\ref{fig:spread}-(b) and -(d), respectively, effectively mitigate the Doppler spread caused by fractional Doppler in the channel.
This mitigation is attributed to the receiver windowing in RW-OTFS, which reduces OTFS sidelobes \cite{b11}, a property similarly achieved by GF-OTFS through global filtering.
In contrast, DR-UFMC does not effectively reduce Doppler spread, as UFMC operates along the delay dimension rather than the Doppler dimension.

\begin{figure}[t!]
\centering
\includegraphics[scale=0.31]{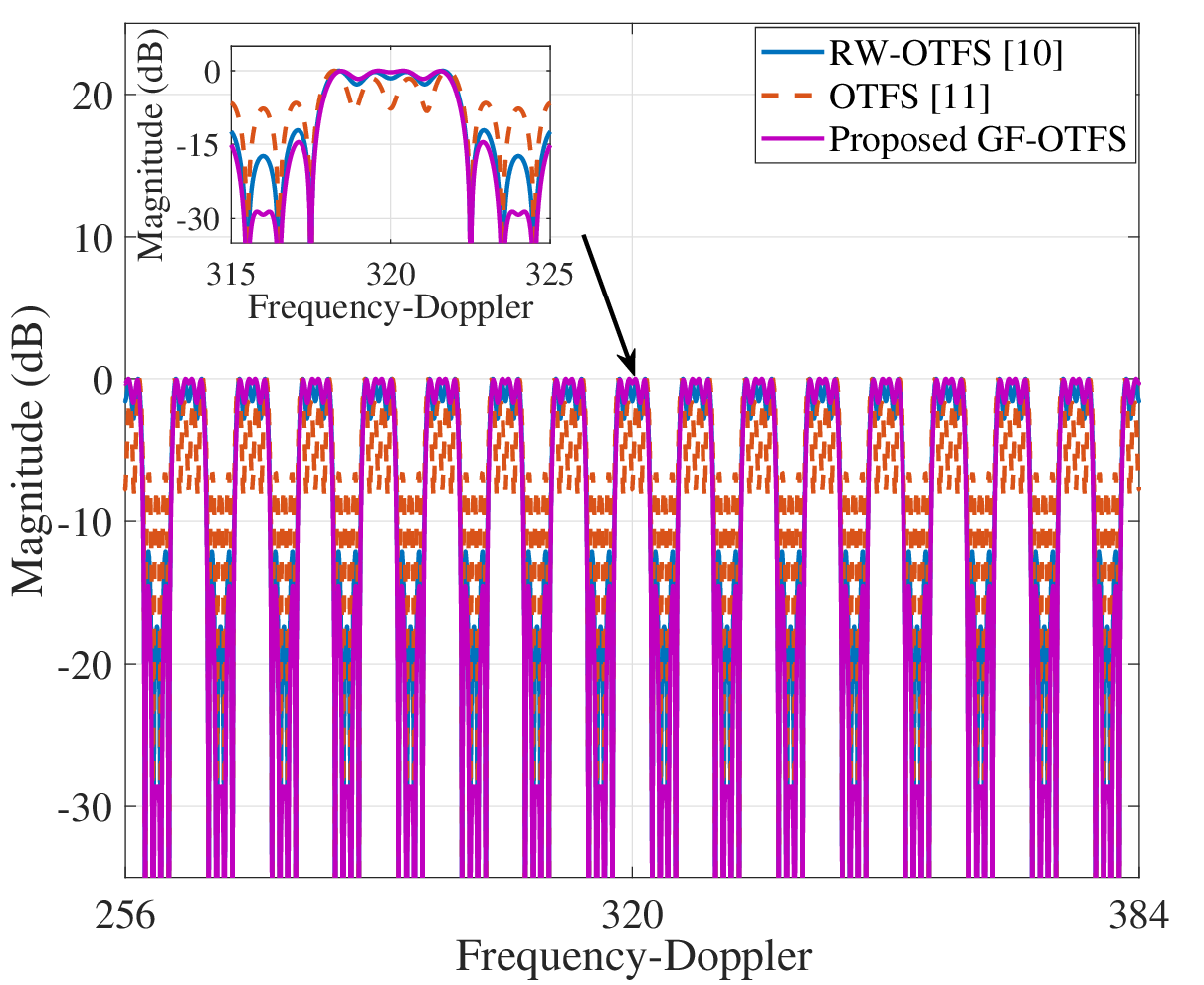} 
\vspace{-0.2cm}
\caption{Sidelobes of individual subbands for the modulation techniques with $M\!\!=\!64$ and $N\!\!=\!8$ within the frequency-Doppler domain spanning $[0,\!M\!N\!-\!1]$.}
\label{fig:sidelobe}
\vspace{-0.2cm}
\end{figure}

Fig.~\ref{fig:sidelobe} shows the spectrum of the transmitted signal for OTFS, RW-OTFS, and GF-OTFS, with only half of the subbands activated.
In this scenario, each subband of GF-OTFS comprises four consecutive frequency-Doppler bins, i.e., $N_{\rm{sc}}^{\rm{rb}}\!=\!4$.
For RW-OTFS, a global window is applied at the transmitter to assess its impact on sidelobe suppression.
RW-OTFS uses a global time-domain window, which can be viewed in the frequency-Doppler domain as filtering each individual frequency-Doppler bin.
Fig. 7 shows the effectiveness of subband filtering in mitigating sidelobes in RW-OTFS and GF-OTFS compared to OTFS.
This mitigation reduces interference and enhances the system's robustness against Doppler effects in LTV channels.
However, GF-OTFS provides improved IDI mitigation by filtering a subband comprising multiple frequency-Doppler bins, whereas RW-OTFS filters each frequency-Doppler bin individually.
It is worth noting that filtering along the delay dimension cannot mitigate the sidelobes caused by Doppler spread and only introduces a frequency shift.
Thus, we do not include the result for DR-UFMC in Fig.~\ref{fig:sidelobe}.

\begin{figure}[!t]
\centering
\includegraphics[scale=0.31]{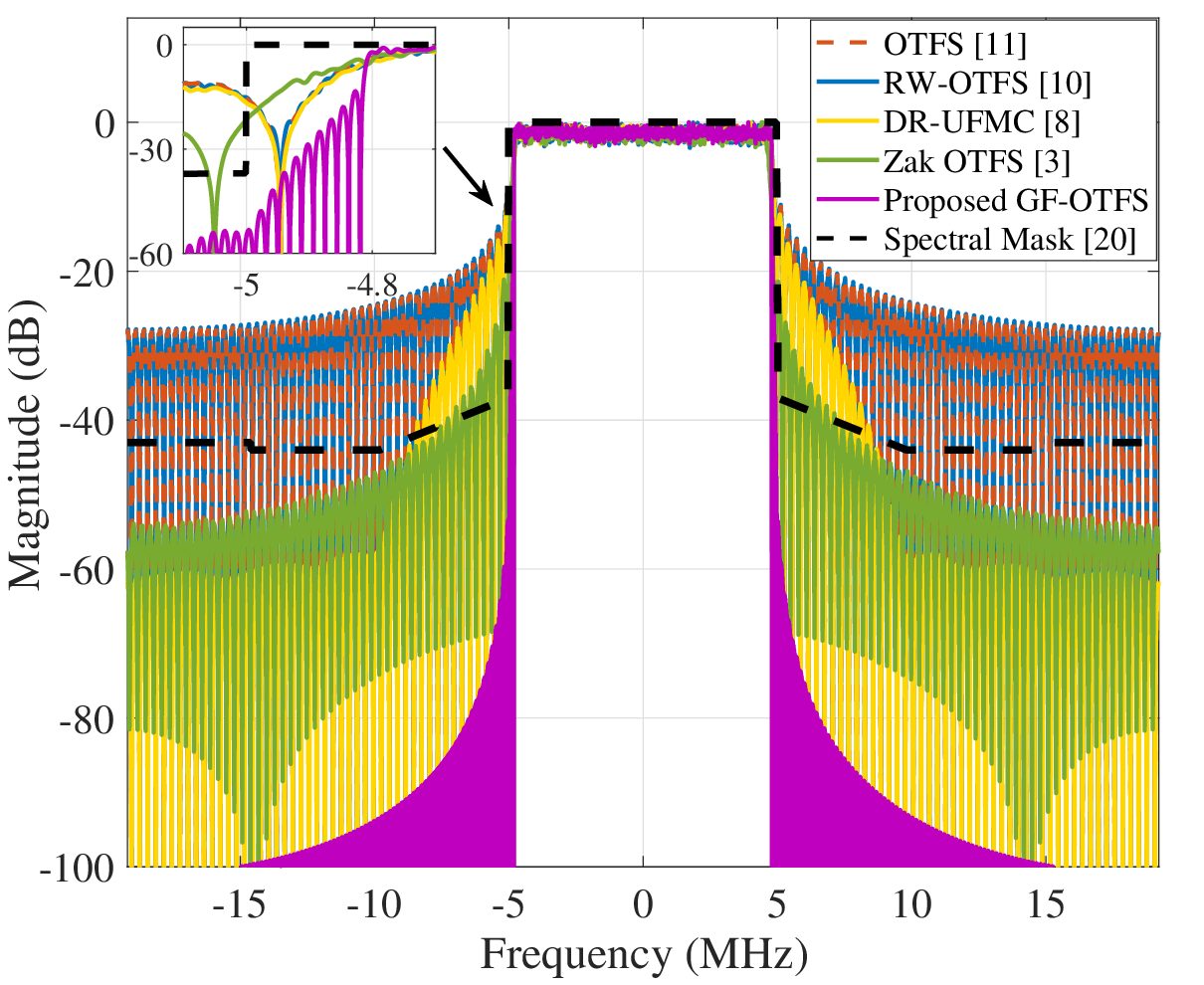} 
\vspace{-0.2cm}
\caption{PSD comparison of different multiplexing techniques for $M=64$, $N=8$, and $B=10~\rm{MHz}$.}
\label{fig:PSD}
\vspace{-0.45cm}
\end{figure}

Fig.~\ref{fig:PSD} compares the power spectral density (PSD) of the different multiplexing techniques studied in this paper. As shown, RW-OTFS exhibits OOB emissions similar to OTFS, indicating that applying a global window does not suppress emissions outside the desired bandwidth.
This occurs because windowing in the time domain is equivalent to circularly convolving the signal spectrum with the transformed window function in the frequency dimension.
The resulting tails of a linear convolution contribute to OOB emissions reduction by providing smooth transitions at the edges of the signal.
However, circular convolution folds these tails back onto the signal itself, causing abrupt changes at the edges and leaving the shape of the OOB emissions unchanged.
Thus, Zak OTFS, ODDM, DR-UFMC, and GF-OTFS, which utilize linear pulse shaping, effectively suppress OOB emissions.
As previously reported in the literature \cite{b20}, Zak OTFS and ODDM exhibit similar OOB emissions, hence, only Zak OTFS is presented in Fig.~\ref{fig:PSD}.
The mitigation of OOB emissions in GF-OTFS and DR-UFMC is attributed to the subband filtering, which attenuates sidelobes across the entire frequency domain.
However, the proposed GF-OTFS provides a significant advantage over DR-UFMC by achieving a much shorter transition band. The shorter transition band minimizes interference between adjacent subbands, making GF-OTFS particularly suitable for operation in a multiuser shared spectrum environment where spectral efficiency is critical.
Additionally, as shown in Fig.~\ref{fig:PSD}, the OOB emissions of the proposed GF-OTFS lie below the spectral mask defined by the third generation partnership project (3GPP) standard for 5G-NR \cite{b13}. This compliance demonstrates that the proposed GF-OTFS multiplexing technique not only improves spectral containment but also adheres to stringent regulatory standards, further reinforcing its suitability for next-generation wireless communication systems.

 \begin{figure}[!t]
\centering
\includegraphics[scale=0.31]{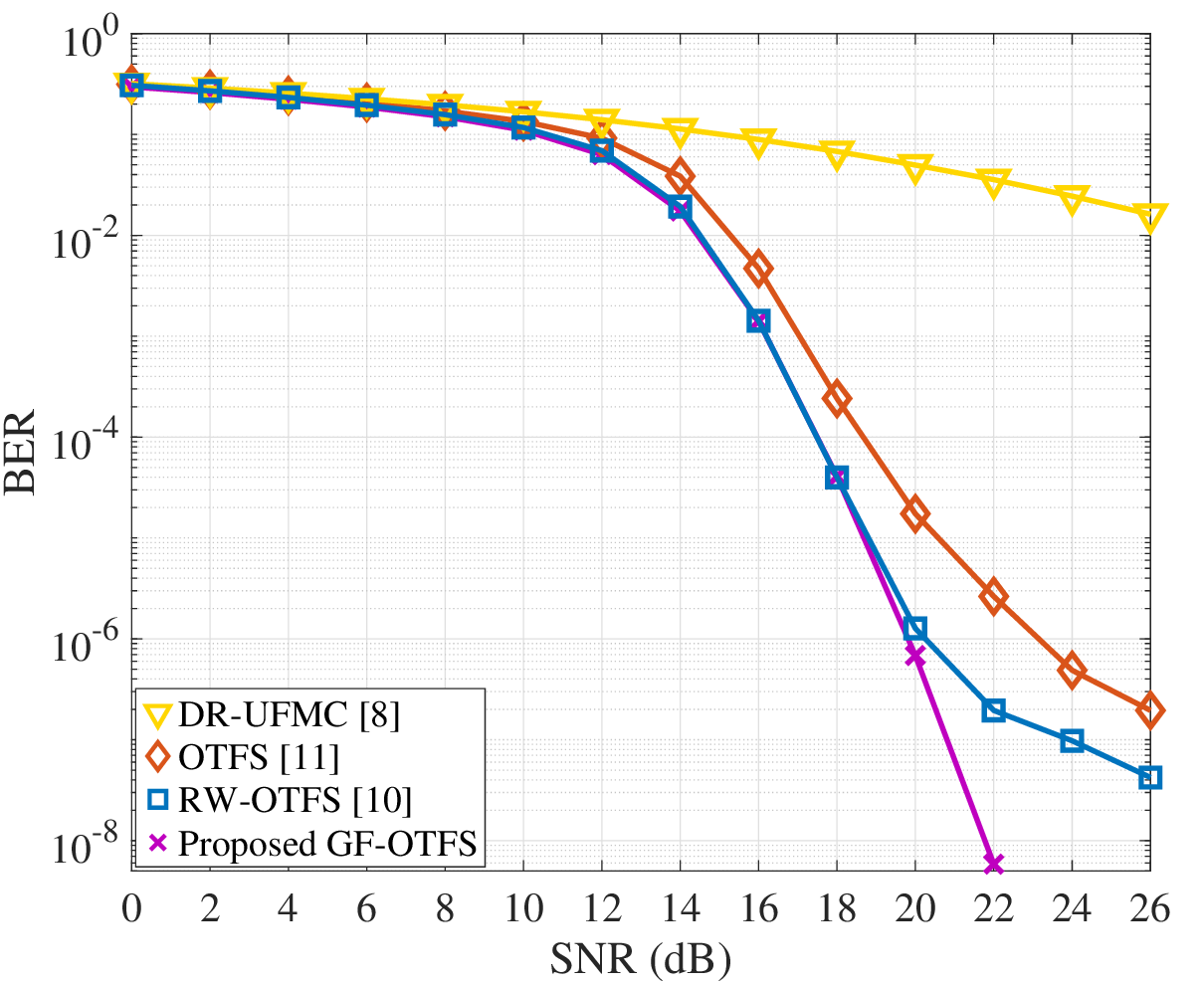} 
\vspace{-0.1cm}
\caption{BER performance of different multiplexing techniques for $16$-QAM, $M=64$, $N=8$, and $B=1.92~\rm{MHz}$.}
\label{fig:BER}
\vspace{-0.3cm}
\end{figure}

The BER performance of the different multiplexing techniques is presented in Fig.~\ref{fig:BER}.
An advanced detector utilizing least squares minimum residual with interference cancellation (LSMR-IC), \cite{b12}, is used in this simulation.
As shown in Fig.~\ref{fig:BER}, our proposed GF-OTFS technique improves the BER performance compared to other multiplexing techniques discussed in this work.
Specifically, GF-OTFS achieves up to $5~\rm{dB}$ better BER performance than RW-OTFS at high SNRs. Moreover, the proposed multiplexing technique significantly outperforms OTFS and DR-UFMC.
It is noteworthy that the BER performance of DR-UFMC is inferior to that of OTFS, as the overlap-and-add process introduces interference between different Doppler bins, which cannot be accurately recovered at the receiver.
Additionally, DR-UFMC utilizes the UFMC modulator along the delay dimension, which does not effectively address the Doppler effect that leads to system performance degradation.
Importantly, with ideal pulse shaping across the delay dimension, Zak OTFS and ODDM achieve the same BER performance as OTFS.

\section{Conclusion}
\label{sec:conclusion}
In this paper, we proposed a novel modulation technique that integrates SC-FDMA-based delay-Doppler representation with UFMC modulation.
Our proposed technique effectively addresses the challenges posed by Doppler shifts in fast time-varying channels, that makes it highly suitable for high-mobility communication scenarios.
The proposed GF-OTFS achieves improved spectral containment and improved robustness against Doppler-induced interference.
Through rigorous mathematical analysis and extensive simulations, we demonstrate that the GF-OTFS approach achieves superior BER performance compared to current techniques while significantly reducing OOB emissions.


\section*{Acknowledgment}

This publication has emanated from research conducted with the financial support of Research Ireland under Grant number 18/CRT/6222, 19/FFP/7005(T) and 21/US/3757. For the purpose of Open Access, the author has applied a CC BY public copyright licence to any Author Accepted Manuscript version arising from this submission.

\bibliographystyle{IEEEtran} 
\bibliography{IEEEabrv,references}

\end{document}